\documentclass[fleqn]{2023SCGE}
\begin{document}

\ensubject{subject}

\ArticleType{Article}
\SpecialTopic{SPECIAL TOPIC: }
\Year{}
\Month{}
\Vol{}
\No{}
\DOI{??}
\ArtNo{000000}
%\ReceiveDate{November 11, 2025}
%\AcceptDate{December 27, 2025}

\title{Nuclear modification of heavy flavor decayed dielectrons in relativistic heavy-ion collisions}{Nuclear modification of heavy flavor decayed dielectrons in relativistic heavy-ion collisions}

\author[1]{Lejing Zhang}{}
\author[2]{Wen-Jing Xing}{{wenjing.xing@usc.edu.cn}}
\author[1]{Shanshan Cao}{{shanshan.cao@sdu.edu.cn}}
\author[3]{Guang-You Qin}{{guangyou.qin@ccnu.edu.cn}}

\AuthorMark{Zhang L J}

\AuthorCitation{Zhang L J, Xing W J, Cao S S, Qin G Y}

\address[1]{Institute of Frontier and Interdisciplinary Science, Shandong University, Qingdao, Shandong 266237, China}
\address[2]{School of Nuclear Science and Technology, University of South China, Hengyang, Hunan 421001, China}
\address[3]{Institute of Particle Physics and Key Laboratory of Quark and Lepton Physics (MOE),\\
Central China Normal University, Wuhan, 430079, China}

\abstract{Dielectrons from heavy flavor hadron decays not only constitute a crucial background to their thermal spectrum in high-energy nuclear collisions, from which the temperature of the quark-gluon plasma (QGP) is extracted, but also provide a valuable probe of heavy quark interactions with the QGP. Using a linear Boltzmann transport (LBT) model to describe heavy quark evolution inside the QGP and a hybrid fragmentation-coalescence model for their hadronization, we find heavy quark energy loss softens the invariant mass spectrum of their decayed dielectrons and yields a higher value of the extracted QGP temperature, while coalescence hardens the spectrum and yields a lower value. Taking into account full medium effects leads to higher values of the extracted temperature than using vacuum baselines of heavy flavor decayed dielectrons in analyzing the experimental data. In addition, we find the angular correlations between dielectron pairs are sensitive to heavy quark interactions with the QGP: the radial flow of the QGP enhances the near-side correlations, and scatterings between heavy quarks and the QGP broaden the away-side correlations, with elastic and string interactions playing a dominant role. 
}

\keywords{relativistic heavy-ion collisions, heavy quarks, dielectrons}
\PACS{47.55.nb, 47.20.Ky, 47.11.Fg}

\maketitle

\begin{multicols}{2}
\section{Introduction}
\label{section1}

Lattice Quantum Chromodynamics (QCD)~\cite{Borsanyi:2010cj,HotQCD:2014kol} predicts that at sufficiently high temperature, nuclear matter exists in a color deconfined state, known as quark-gluon plasma (QGP). The only way of
achieving such high temperature in the laboratory is to collide beams of ultrarelativistic heavy ions at the Relativistic Heavy
\Authorfootnote
Ion Collider (RHIC) and the Large Hadron Collider (LHC). It is now generally accepted that QGP is formed in these high-energy nuclear collisions, as revealed by the collective flow of hadrons and the significantly suppressed yields of energetic particles, or jet quenching, in these collisions~\cite{Gyulassy:2004zy,Jacobs:2004qv,Muller:2012zq,Elfner:2022iae,Li:2024uzk,Chen:2024aom,Shou:2024uga,Lu:2025qyf}.

While there are many indirect ways of inferring the temperature of the QGP, such as comparing hydrodynamic simulations of the QGP with the low transverse momentum ($p_\mathrm{T}$) hadron data~\cite{Bernhard:2019bmu,JETSCAPE:2020shq,Nijs:2020roc}, comparing jet quenching calculations with the high $p_\mathrm{T}$ data~\cite{JET:2013cls,JETSCAPE:2021ehl,Armesto:2009zi,Andres:2016iys,Faraday:2025pto}, a direct measurement of the QGP temperature has remained a challenge for years. One may extract the chemical or kinetic freeze-out temperature by comparing thermal model calculations with the yields and $p_\mathrm{T}$ spectra of different hadrons~\cite{Andronic:2017pug,Chen:2020zuw,STAR:2017sal}, but this is the temperature when stable hadrons form instead of that of the color deconfined QGP state. One may constrain the QGP temperature based on the sequential melting of heavy quarkonium states inside the QGP~\cite{Matsui:1986dk,Satz:2005hx,CMS:2012gvv}, but the resolution is limited by the dissociation temperature gap between two quarkonium states, and can also be affected by regeneration of quarkonia inside the QGP~\cite{Du:2017qkv,Zhao:2017yan}. The spectra of direct photons have been used to extract the average temperature of the QGP~\cite{Sahlmuller:2015mfq}, but this method suffers from the blue shift effect due to the fast expansion of the QGP. The best candidate of the QGP thermometer is the invariant mass spectrum of thermal dileptons ($e^+e^-$ or ${\mu}^+{\mu}^-$ pairs) produced from virtual photons emitted by the QGP. This observable benefits from both its Lorentz invariance property that protects it from the blue shift effect, and negligible nuclear modification of virtual photons and dileptons once they form~\cite{STAR:2024bpc,vanHees:2006ng,Rapp:2014hha}. Considerable efforts have been devoted to extracting the QGP temperature using the dilepton spectrum~\cite{NA60:2008dcb,STAR:2012dzw,STAR:2013pwb}, and a final success has been recently achieved in Au+Au collisions at center-of-mass energies per nucleon pair of $\sqrt{s_\mathrm{NN}}=54.4$~GeV and 27~GeV \cite{STAR:2024bpc}. Preliminary results for Isobar (Ru+Ru and Zr+Zr) collisions at $\sqrt{s_\mathrm{NN}}=200$~GeV have also been reported at the Quark Matter 2025 conference~\cite{Luo:2025QM}.

In reality, signals of thermal dileptons are embedded in a non-thermal background of dileptons, known as the cocktail sum~\cite{Rapp:2013nxa}. This includes decays from $\pi^0$, $\eta$, $\omega$, $\eta'$, and $\phi$ mesons, which dominate dilepton production in the low invariant mass region ($M_{ee}<M_{\phi}\simeq1.02~\mathrm{GeV}$), and the Drell-Yan process and decays from $J/\psi$, which dominate in the high invariant mass region ($M_{ee}>M_{J/\psi}\simeq3.10~\mathrm{GeV}$). The intermediate invariant mass region ($M_{\phi}<M_{ee}<M_{J/\psi}$) is relatively clean for thermal dileptons, where the non-thermal background is mainly contributed by the Drell-Yan process and decays from open heavy flavor hadrons. This is where the extraction of the QGP temperature is usually conducted. Although nuclear modification of heavy flavor hadrons has been extensively explored~\cite{Andronic:2015wma,Cao:2018ews,Xu:2018gux,Dong:2019byy,He:2022ywp,Zhao:2023nrz}, there are rare studies on how this nuclear modification affects the dilepton spectra decayed from heavy flavor hadrons~\cite{Xu:2013uza}. So far, in experiments, the background of heavy flavor decayed dileptons is simulated using Pythia~\cite{Sjostrand:2006za,Bierlich:2022pfr}, an event generator for proton-proton ($p+p$) collisions. The QGP effects in nucleus-nucleus (A$+$A) collisions are only roughly estimated by either randomizing the momentum directions of leptons or reducing the energies of leptons, mimicking the momentum broadening or energy loss of heavy quarks inside the QGP~\cite{STAR:2024bpc}. One major goal of this work is quantitatively exploring how nuclear modification of heavy flavor hadrons, including energy loss of heavy quarks and their in-medium hadronization, affects the spectra of their decayed dielectrons, and evaluating the systematic uncertainties it introduces to the QGP temperature extracted from thermal dielectrons.

Apart from modifying the invariant mass spectra of dileptons, scatterings between heavy quarks and the QGP can also alter the angular correlations between dilepton pairs. It has been proposed that the angular correlations between $D\overline{D}$ meson pairs can be used to probe the early stage thermalization of light partons in heavy-ion collisions~\cite{Zhu:2006er}. These correlations are also suggested as promising observables for resolving the individual strengths of elastic and inelastic scatterings endured by heavy quarks inside the QGP~\cite{Nahrgang:2013saa,Cao:2015cba}. Although significant efforts have been made to measure the $D\overline{D}$ correlations in $p+p$ collisions~\cite{LHCb:2012aiv,Ma:2017ybx}, extending such measurements to A$+$A collisions is still considered challenging at both RHIC and the LHC. Alternative options could be studying the correlations between $D$ mesons and charged particles, or correlations between heavy flavor decayed leptons~\cite{ALICE:2019oyn,ATLAS:2023vms,Thomas:2024cso}. Considering that nuclear modification of bottom quarks is relatively weak and its impact on the angular correlations between their decayed dileptons can be hardly seen in the experimental data~\cite{ATLAS:2023vms}, we will focus on the correlations between charm decayed dielectrons in this work. This could serve as a potential new observable of open heavy flavor, in addition to those of single inclusive hadrons~\cite{STAR:2014wif,ALICE:2015vxz,CMS:2017uoy,CMS:2017qjw,Tang:2020ame,Zhang:2025pqu} and their decayed leptons~\cite{PHENIX:2005nhb,ATLAS:2018ofq,ALICE:2016uid,ATLAS:2021xtw}, and provide additional insights into the dynamics of heavy quark interactions with the QGP.

\section{Heavy flavor evolution in heavy-ion collisions}
\label{sec:2}

Due to their large masses, thermal production of heavy quarks from the QGP is neglected. We use Pythia~\cite{Sjostrand:2006za,Bierlich:2022pfr} to generate heavy quarks from the initial hard nucleon-nucleon scatterings. The positions of these hard scatterings in nucleus-nucleus collisions are sampled using the Monte-Carlo Glauber model. 

Interactions between heavy quarks and the QGP medium are described using the linear Boltzmann transport (LBT) model~\cite{Luo:2023nsi,Xing:2021xwc}, with both elastic and inelastic scatterings, and both perturbative and non-perturbative interactions included. In LBT, the phase space distribution of heavy quarks, $f_a(x_a,p_a)$, evolves according to the Boltzmann equation as
\begin{align}
	\label{eq:boltzmann1}
	p_a \cdot\partial f_a(x_a,p_a)=E_a (\mathcal{C}_\mathrm{el}+\mathcal{C}_\mathrm{inel}).
\end{align}
From the collision integral contributed by elastic scatterings, $\mathcal{C}_\mathrm{el}$, one may extract the elastic scattering rate (number of elastic scatterings per unit time) as~\cite{Cao:2016gvr,Xing:2021xwc},
\begin{align}
	\label{eq:gamma_el}
	\Gamma^\mathrm{el}_{a} & (\vec{p}_a, T) = \sum_{b,(cd)}\frac{\gamma_b}{2E_a}\int \frac{d^3 p_b}{(2\pi)^3 2E_b} \frac{d^3 p_c}{(2\pi)^3 2E_c} \frac{d^3 p_d}{(2\pi)^3 2E_d}\nonumber
	\\ \times\,& f_b (\vec{p}_b, T) \left[1 \pm f_c (\vec{p}_c, T)\right]
	\left[1 \pm f_d (\vec{p}_d, T)\right]\nonumber
	\\ \times\,&\theta \left[s - (m_a + \mu_d)^2\right] \nonumber 
	\\ \times\,& (2\pi)^4 \delta^{(4)}(p_a + p_b - p_c -p_d) \left|\mathcal{M}_{ab \rightarrow cd}\right|^2.
\end{align}
The summation is over all possible $ab \rightarrow cd$ scattering channels, with $b$ a medium constituent parton, $c$ and $d$ the final states of $a$ and $b$, respectively. In addition, $\gamma_b$ is the spin-color degrees of freedom of $b$, $f_{b,c,d}$ takes the form of thermal (Bose or Fermi) distribution at medium temperature $T$, $s$ is the Mandelstam variable, $m_a$ is the heavy quark mass, and $\mu_d$ denotes the Debye screening mass.

\begin{figure}[H]
	\centering
	\includegraphics[scale=0.22]{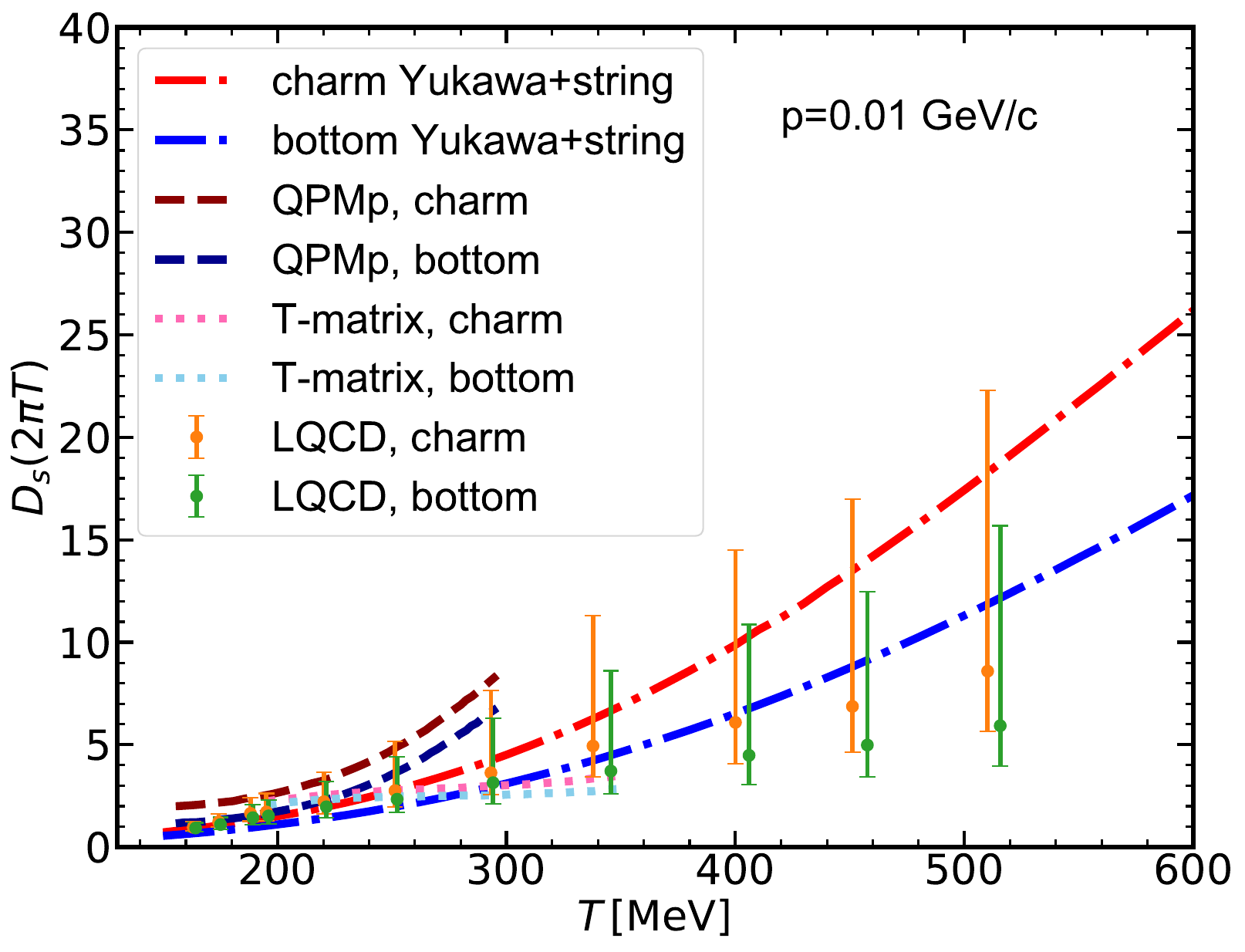}
	\caption{(Color online) Temperature dependences of the spatial diffusion coefficients of charm and bottom quarks in the LBT model with Yukawa and string interactions, compared to results from the lattice QCD calculation~\cite{HotQCD:2025fbd}, the QPMp model~\cite{Sambataro:2024mkr}, and the non-perturbative $T$-matrix model~\cite{Liu:2016ysz,Tang:2023tkm}.} 
	\label{fig:Ds}
\end{figure}

In this work, we use the version of the model~\cite{Xing:2021xwc}, in which both perturbative and non-perturbative interactions between heavy quarks and the QGP are incorporated in the scattering matrices as
\begin{align}
	\label{eq:Matrix_cq}
	i \mathcal{M}  =\,& i\mathcal{M_\mathrm{Y}} + i\mathcal{M_\mathrm{S}}\nonumber
	\\=\,& \overline{u}(p') \gamma^{\mu} u(p) V_\mathrm{Y}(\vec{q}) \overline{u}(k') \gamma_{\mu} u(k)\nonumber
	\\&+ \overline{u}(p') u(p) V_\mathrm{S}(\vec{q}) \overline{u}(k') u(k),
\end{align}
where a vector interaction Yukawa term ($\mathcal{M}_\mathrm{Y}$) is used to model perturbative scatterings~\cite{Combridge:1978kx} and a scalar interaction string term ($\mathcal{M}_\mathrm{S}$) is assumed to model non-perturbative scatterings~\cite{Riek:2010fk}. There is no interference between vector and scalar interactions, and thus the total scattering rate is contributed by these two terms independently. This enables investigation of dynamical effects of perturbative and non-perturbative interactions individually within LBT, as will be discussed later in Sec.~\ref{sec:4}. In particular, the Yukawa term dominates at high temperature and large momentum, whereas the string term becomes prominent at low temperature and small momentum. The effective propagators, ${V}_\mathrm{Y}$ and ${V}_\mathrm{S}$, can be obtained from the Fourier transformation of a Cornell type potential and read,
\begin{align}
	V_\mathrm{Y}(\vec{q}) = - \frac{4\pi \alpha_\mathrm{s} C_F}{m_d^2+|\vec{q}|^2}, \quad\quad V_\mathrm{S}(\vec{q}) = - \frac{8\pi \sigma}{(m_s^2+|\vec{q}|^2)^2}.
	\label{eq:potential}
\end{align}
Here, $\vec{q}$ denotes the momentum exchange in elastic scatterings, and $C_F=4/3$ is the color factor. Detailed expressions of the scattering matrices can be found in Ref.~\cite{Xing:2021xwc}, in which the following model parameters are determined by the nuclear modification data of heavy flavor hadrons at RHIC and the LHC: $\alpha_\mathrm{s}=0.27$ is the strong coupling coefficient, $\sigma=0.45~\mathrm{GeV}^2$ represents the strength of the string term, and $m_d=a+bT$ and $m_s=\sqrt{a_s+b_s T}$ are the screening masses in the Yukawa and string terms, respectively, with $a=0.20$~GeV, $b=2.0$, $a_s=0$, and $b_s=0.10$~GeV. These parameters also provide a satisfactory description of the nuclear modification factors of heavy flavor decay products~\cite{Dang:2023tmb,Xing:2024qcr} and $B_\mathrm{c}$ mesons~\cite{Zhang:2025cvk}. Figure~\ref{fig:Ds} presents the spatial diffusion coefficients of heavy quarks obtained using the parameter set above in the zero momentum limit. The results are consistent with those from the latest lattice QCD calculation over an extended temperature range~\cite{HotQCD:2025fbd}, as well as the QPMp framework that includes the lattice-constrained quark susceptibilities~\cite{Sambataro:2024mkr}, and the non-perturbative $T$-matrix calculation~\cite{Liu:2016ysz,Tang:2023tkm}.

The inelastic part of the collision integral in Eq.~(\ref{eq:boltzmann1}), $\mathcal{C}_\mathrm{inel}$, can be related to the rate of medium-induced gluon emissions from heavy quarks. We take the gluon spectrum from the higher-twist energy loss calculation~\cite{Guo:2000nz,Zhang:2003wk,Majumder:2009ge}, in which the rate of gluon emission is proportional to the jet transport coefficient $\hat{q}_a$. This $\hat{q}_a$  characterizes the transverse momentum broadening square of a hard parton per unit time arising from elastic scatterings, and can be obtained by integrating Eq.~(\ref{eq:gamma_el}) with an additional weight factor $\left[ \vec{p}_c - (\vec{p}_c \cdot \hat{p}_a) \, \hat{p}_a \right]^2$. Numerical implementation of these elastic and inelastic scatterings is detailed in Ref.~\cite{Luo:2023nsi}.

The QGP medium is simulated using the (3+1)-dimensional viscous hydrodynamic model CLVisc~\cite{Pang:2012he,Pang:2018zzo,Wu:2018cpc,Wu:2021fjf}, which is constrained by the soft hadron data. Before the hydrodynamic evolution of the medium starts ($\tau_0=0.6$~fm), heavy quarks are assumed to stream freely from their production vertices. After that, the hydrodynamic simulation provides the local temperature and fluid velocity of the QGP medium, based on which one can calculate the collision integrals in Eq.~(\ref{eq:boltzmann1}) and therefore simulate the phase space evolution of heavy quarks. Note that by keeping the temperature information but manually setting fluid velocity to zero in simulation, one may investigate how the radial flow of the QGP affects the heavy flavor observables, as will be discussed later in Sec.~\ref{sec:4}. On the hypersurface of medium temperature at $T_\mathrm{pc}=165$~MeV, we convert heavy quarks into heavy flavor hadrons using a hybrid fragmentation-coalescence model~\cite{Cao:2019iqs}, which is constrained by the RHIC and LHC data on the heavy flavor hadron chemistry. In the end, we use Pythia~\cite{Sjostrand:2006za} to simulate the decays of heavy flavor hadrons into electrons.

In hadronization, the momentum distribution of the final state hadrons produced from heavy-light quark coalescence is given by
	\begin{equation}
	\label{eq:coalescence}
		f_h(\vec{p}^{\,'}_h) = \int \left[ \prod_i d\vec{p}_i f_i(\vec{p}_i) \right] W(\{\vec{p}_i\}) \delta\left(\vec{p}^{\,'}_h - \sum_i \vec{p}_i\right), 
	\end{equation}
with $\vec{p}^{\,'}_h$ denoting the three-momentum of the final state hadron, and $\vec{p}_i$ representing the three-momentum of each constituent quark. The Wigner function $W$ quantifies the probability for quarks to combine into a hadron. It is defined as the wavefunction overlap between the free quark state and the hadron bound state.
Assuming a two-body quark system is described by a harmonic oscillator model, the Wigner functions for $s$- and $p$-wave meson formations are given by
	\begin{align}
		W_s &= g_h \frac{(2\sqrt{\pi}\sigma)^3}{V} e^{-\sigma^2 k^2},  \\
		W_p &= g_h \frac{(2\sqrt{\pi}\sigma)^3}{V} \frac{2}{3} \sigma^2 k^2 e^{-\sigma^2 k^2}. 
	\end{align}
Here, $g_h$ is the statistical factor accounting for spin and color degrees of freedom, $k$ denotes the relative three-momentum between the constituent quarks in the rest frame of the heavy flavor meson, $\omega$ is the harmonic oscillator frequency related to the meson radius~\cite{Cao:2019iqs}, $\mu$ is the reduced mass of the two constituent quarks, and $\sigma = 1/\sqrt{\mu \omega}$. When deriving the Wigner function for heavy flavor baryons, we need to first combine two of the constituent quarks and then combine their center of mass with the third constituent quark. By setting the heavy quark distribution as a Dirac delta function and the light quark distribution as a thermal distribution on the right hand side of Eq.~(\ref{eq:coalescence}), its left hand side gives the coalescence probability of a heavy quark with a given momentum into a particular type of hadron. Based on such probability, if a heavy quark does not turn into any hadron via coalescence, we use Pythia to simulate its hadronization via fragmentation.

The independent fragmentation module of Pythia is applied for the fragmentation process, in which the following Peterson fragmentation function is used
\begin{equation}
\label{eq:peterson}
	D_{Q}^{H}(z)\propto\frac{1}{z\left[1-(1 / z)-\epsilon_{Q} /(1-z)\right]^{2}},
\end{equation}
in which $z = p_H / p_Q$ represents the fractional momentum taken by a heavy flavor hadron relative to its parent heavy quark, and $\epsilon_{Q}$ is a model parameter. If not otherwise specified, we keep their default values, $\epsilon_{c}=0.05$ for charm quarks and $\epsilon_{b}=0.005$ for bottom quarks, in our work. However, in Sec.~\ref{sec:3}, $\epsilon_{c}$ is temporarily re-adjusted to reproduce the invariant mass spectra of the charm decayed dielectrons used by the STAR Collaboration in their data analysis, which is obtained using the string fragmentation module of Pythia. Effects of varying this parameter will be discussed.

\section{Invariant mass spectra of heavy quark decayed dielectrons}
\label{sec:3}

\begin{figure*}[tbp]
	\centering
	\includegraphics[width=0.75\textwidth]{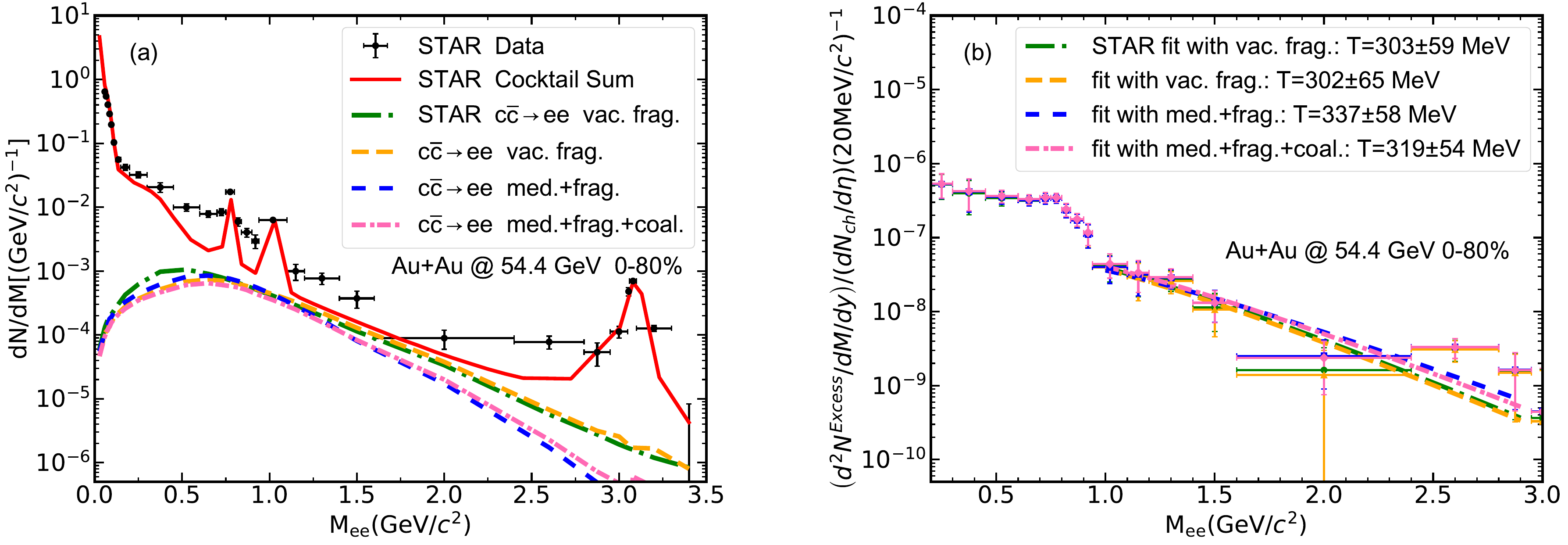}
	\caption{(Color online) The invariant mass spectra of dielectrons in 0-80\% Au+Au collisions at $\sqrt{s_\mathrm{NN}} = 54.4$~GeV. Panel (a) compares between different model calculations of the contributions from charm quark decay, the STAR data on the total yield and the cocktail sum used by STAR in analyzing the data~\cite{STAR:2024bpc}. Panel (b) compares between different fittings (lines) of the QGP temperature to the invariant mass spectra of thermal dielectrons (cross symbols), obtained by subtracting the STAR data on the total yield by different cocktail sums that involve different estimations of the charm decayed dielectrons. 
}
	\label{fig:54.4}
\end{figure*}

We first explore nuclear modification of the invariant mass spectra of heavy quark decayed dielectrons and its impact on the QGP temperature extracted from the dielectron spectra.
\begin{table}[H]
	\centering
	\footnotesize
	\begin{threeparttable}
		\caption{Parameter settings of \textsc{Pythia}~6 simulation.}
		\label{tab:pythia}
		\doublerulesep 0.1pt \tabcolsep 13pt
		\renewcommand{\arraystretch}{1.2}
		\begin{tabular}{ccp{3.6cm}}
			\toprule
			\textbf{parameter} & \textbf{value} & \textbf{function} \\
			\hline
			MSEL & 1 & Minibias trigger of Pythia \\
			PARP(91) & 1.0 & Width of the Gaussian primordial $k_\mathrm{T}$ distribution inside hadron \\[1ex]
			PARP(67) & 1.0 & The $Q^2$ scale of the hard scattering is multiplied by this value \\
			\bottomrule
		\end{tabular}
	\end{threeparttable}
\end{table}

In order to directly compare to the experimental results, we use the same Pythia 6 simulation for heavy quark production as implemented by the STAR Collaboration~\cite{Wang2023}. Key parameter tunings and their physical meanings are listed in Tab.~\ref{tab:pythia}. These heavy quarks then scatter through the QGP medium within the LBT model, turn into heavy flavor hadrons within the hybrid fragmentation-coalescence model, and decay into electrons within Pythia 6. To improve simulation efficiency, non-semileptonic decay channels of heavy flavor hadrons are switched off. Following the STAR analysis, we only consider dielectrons decayed from $D$ mesons and $\Lambda_c$ baryons for charm quarks. For consistency, only dielectrons decayed from $B$ mesons and $\Lambda_b$ baryons are considered for bottom quarks. The following normalization formula from Ref.~\cite{Wang2023} is employed to obtain the final spectra of the heavy quark decayed dielectrons: 
\begin{align}
	\begin{split}
		\frac{1}{N_\mathrm{mb}} \frac{d N}{d M_{ee}} &= \frac{1}{N_\mathrm{evt}} \left(\frac{d N}{d M_{ee}}\right)_{pp} \frac{\sigma_{Q\bar{Q}}}{\sigma_\mathrm{mb}} N_\mathrm{Bin} \\
		&\times B R\left(Q/\bar{Q} \rightarrow e^{+}\right) B R\left(Q/\bar{Q} \rightarrow e^{-}\right).
	\end{split}
\end{align}
On the right side, we analyze events that contain at least one heavy quark produced in one nucleon-nucleon binary collision. The number of such events is denoted by ${N_\mathrm{evt}}$, and the number of binary collisions in one nucleus-nucleus collision is denoted by $N_\mathrm{Bin}$. For the cross section of heavy quark production in $p+p$ collisions $\sigma_{Q\bar{Q}}$, we take $\sigma_{c\bar{c}} = 72~\mu\mathrm{b}$ for charm quarks in Au+Au collisions at $\sqrt{s_\mathrm{NN}} = 54.4~\mathrm{GeV}$~\cite{Wang2023}, and ignore dileptons decayed from bottom quarks at this low energy. In Isobar collisions at $\sqrt{s_\mathrm{NN}} = 200~\mathrm{GeV}$, the production cross sections are taken as $\sigma_{c\bar{c}} = 443~\mu\mathrm{b}$ for charm quarks and $\sigma_{b\bar{b}} = 3.7~\mu\mathrm{b}$ for bottom quarks\footnote{Private communication with the STAR Collaboration.}. The inelastic scattering cross section of $p+p$ collisions, $\sigma_\mathrm{mb}$, is taken as $35~\mathrm{mb}$ for $\sqrt{s_\mathrm{NN}} = 54.4~\mathrm{GeV}$ and $42~\mathrm{mb}$ for $\sqrt{s_\mathrm{NN}} = 200~\mathrm{GeV}$. The branching ratios ($BR$) of heavy flavor hadrons decaying into electrons and positrons are taken from the Particle Data Group~\cite{ParticleDataGroup:2022pth}. 

In Fig.~\ref{fig:54.4}(a), we present the invariant mass spectra of charm decayed dielectrons in 0-80\% Au+Au collisions at $\sqrt{s_\mathrm{NN}} = 54.4$~GeV. The STAR Collaboration simulates the dielectron spectra from semileptonic decays of charm quarks in $p+p$ collisions using Pythia 6, and scales them to Au+Au collisions by $N_\mathrm{Bin}$. The hadronization of charm quarks is simulated using the string fragmentation module in Pythia. However, due to the difficulty in tracking the color information of heavy quarks inside the QGP, the independent fragmentation module with the Peterson fragmentation function is applied in our hybrid fragmentation-coalescence model after heavy quarks exit the QGP. For consistency, our $p+p$ baseline (denoted as ``vac. frag.") is also obtained using the same fragmentation routine. As shown in Fig.~\ref{fig:54.4}(a), the $\epsilon_c$ parameter in Eq.~(\ref{eq:peterson}) needs to be adjusted to 0.0003 in order to reproduce the dielectron spectra obtained by STAR (orange vs. green lines). This is smaller than its default value 0.05. Varying $\epsilon_c$ affects other heavy flavor observables, such as the spectrum of $D$ mesons in $p+p$ collisions and their $R_\mathrm{AA}$ in Au+Au collisions. However, since the goal of the present study is to understand the systematic uncertainty of the extracted QGP temperature from medium modification of heavy flavor decayed dielectrons, we choose to tune our $p+p$ baseline of dielectron spectrum to match that used in the STAR analysis, on top of which we apply medium modification constrained by the nuclear modification data of heavy flavor hadrons in our earlier work~\cite{Xing:2021xwc}, where $\epsilon_c=0.05$ was used. Although an inconsistency exists, our approach incorporates realistic medium modification effects in extracting the QGP temperature, and meanwhile enables a controlled comparison with the STAR result. We note that the string fragmentation setup used by STAR in this analysis leads to a consistent $D$ meson spectrum with that from our independent fragmentation setup using $\epsilon_c=0.0003$ in $p+p$ collisions at $\sqrt{s}=200$~GeV, both deviating from the corresponding STAR data ~\cite{STAR:2012nbd}. Impacts of the $\epsilon_c$ parameter on the extracted QGP temperature will be discussed later.  Furthermore, STAR applies acceptance corrections to their Pythia baseline. Although these corrections affect the spectra of charm decayed dielectrons in the low mass region, they can be neglected for the intermediate mass region we focus on. We have confirmed that our $p+p$ baseline is consistent with the uncorrected $p+p$ result of STAR. 

When interactions between heavy quarks and the QGP are taken into account, we observe an apparent softening of the dielectron spectrum (leftward shift towards the low mass region), which can be attributed to the energy loss experienced by charm quarks inside the QGP. The energy loss drives the charm quark pairs towards equilibrium and reduces their relative momenta. Compared to pure fragmentation of medium-modified charm quarks (blue line), including coalescence for their hadronization (pink line) hardens the dielectron spectrum, since coalescence enhances the heavy flavor hadron energy when charm quarks combine with light quarks.

\begin{figure*}[tbp]
	\centering
	\includegraphics[width=0.75\textwidth]{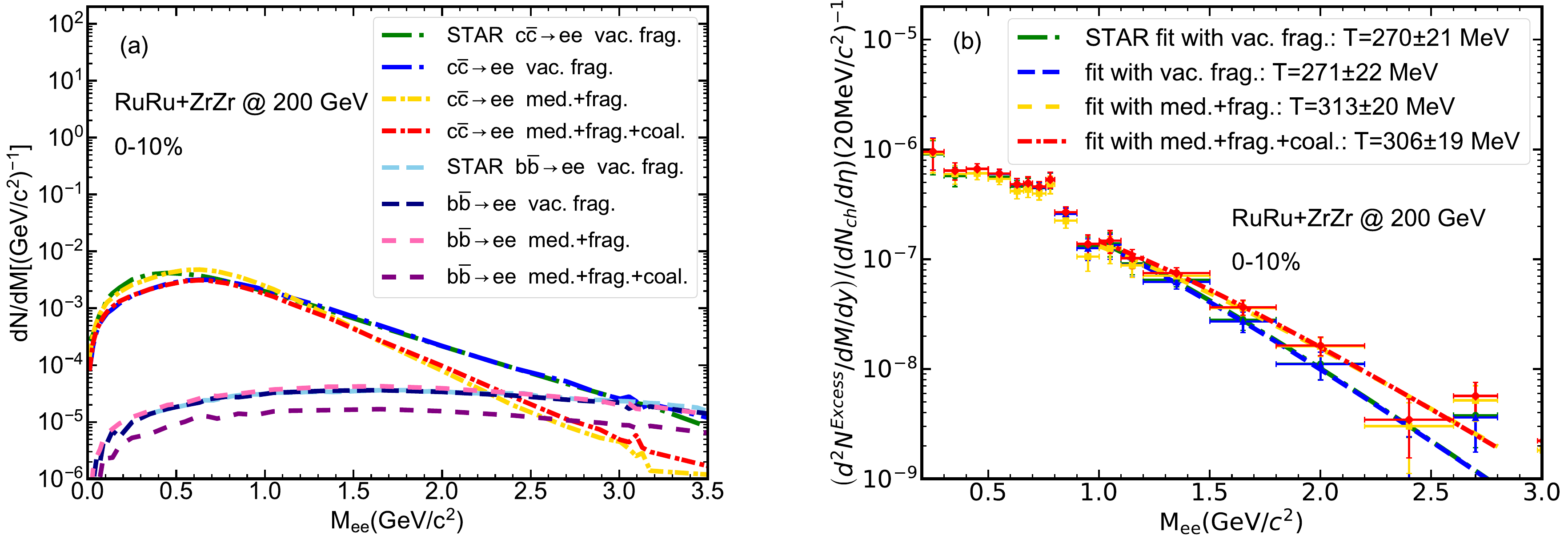}
	\caption{(Color online) The invariant mass spectra of dielectrons in 0-10\% Isobar collisions at $\sqrt{s_\mathrm{NN}} = 200$~GeV. Panel (a) compares between different model calculations of the contributions from charm and bottom quark decays. Panel (b) compares between different fittings (lines) of the QGP temperature to the invariant mass spectra of thermal dielectrons (cross symbols), obtained by subtracting the preliminary STAR data on the total yield (reported at Quark Matter 2025 conference~\cite{Luo:2025QM}) by different cocktail sums that involve different estimations of the heavy quark decayed dielectrons.
}
	\label{fig:200}
\end{figure*}

We replace the STAR estimation of the charm decayed dielectron spectrum by our estimation in the cocktail sum of the spectrum, and then subtract the cocktail sum from the full yield data from STAR. This gives the invariant mass spectrum of thermal dielectrons as shown in Fig.~\ref{fig:54.4}(b).
In the intermediate mass region ($1.0 < M_{ee} < 2.9~\mathrm{GeV}$), we extract the QGP temperature by fitting a Boltzmann function $M^{3/2} \exp(-M/T)$~\cite{STAR:2024bpc,HADES:2019auv,Specht:2010xu} to the spectrum of thermal dielectrons. Although different fitting formulae were used in literature, e.g., $\mathrm{d}^{2} N_{q \bar{q}}^{e e} / (\mathrm{d} M\, \mathrm{d} t) \propto (M T)^{3/2} \exp(-M/T)$~\cite{Rapp:2014hha} and $\mathrm{d}^{2} N_{q \bar{q}}^{e e} / (\mathrm{d} M\, \mathrm{d} t) \propto T^{-9/2} M^{3/2} \exp(-M/T)$~\cite{Zhou:2024yyo}, we have checked that the extracted value of temperature is dominated by the exponential term and is insensitive to different forms of the power term. The values of the extracted QGP temperature are labeled in the figure legend. Our result obtained using the vacuum baseline of charm quarks ($302\pm65$~MeV) agrees with that obtained by STAR ($303\pm59$~MeV). However, when medium modification of heavy quarks is taken into account, followed by pure fragmentation for hadronization, the extracted temperature increases to $337\pm58$~MeV. If coalescence is further included, the extracted temperature decreases to $319\pm54$~MeV. These are consistent with the softening and hardening of the dielectron spectrum from charm quark decay due to energy loss and coalescence, respectively, as shown in Fig.~\ref{fig:54.4}(a). Only statistical uncertainties of the experimental data are taken into account in our fitting. Our study suggests that in 0-80\% Au+Au collisions at $\sqrt{s_\mathrm{NN}} = 54.4$~GeV, heavy quark interactions with the QGP can lead to a systematic uncertainty of about 35~MeV in the temperature extracted from the dielectron spectrum, and the hadronization scheme can lead to another uncertainty of about 18~MeV. It should be emphasized that heavy quark energy loss and heavy-light parton coalescence are not uncertainties of theoretical models of heavy quarks. They are essential ingredients without which models cannot describe the nuclear modification data of heavy flavor hadrons and their decay products. On the other hand, including or not including these ingredients in data analysis leads to different values of the extracted QGP temperature, contributing to the systematic uncertainties in the experimental results. For this reason, we call them ``uncertainties'' in this work from the experimental point of view.

\begin{table}[H]
	\centering
	\footnotesize
	\begin{threeparttable}
		\caption{The extracted values of the QGP temperature (in MeV) for different centrality regions of Au+Au collisions at $\sqrt{s_\mathrm{NN}} = 54.4$~GeV, compared between using charm decayed dielectron spectra with different model setups.}
		\label{tab:54.4_0.0003}
		\setlength{\tabcolsep}{6pt} 
		\renewcommand{\arraystretch}{1.1} 
		\begin{tabular}{cccc}
			\toprule
			centrality & vac. frag. & med.+frag. & med.+frag.+coal.\\
			\midrule
			0-10\% & $347 \pm 148$ & $371 \pm 121$ & $353 \pm 115$ \\
			10-40\% & $263 \pm 61$ & $293 \pm 57$ & $281 \pm 54$ \\
			0-80\% & $302 \pm 65$ & $337 \pm 58$ & $319 \pm 54$ \\
			40-80\% & $286 \pm 73$ & $306 \pm 65$ & $295 \pm 66$ \\
			\bottomrule
		\end{tabular}
	\end{threeparttable}
\end{table}

\begin{table*}[ht!]
	\centering
	\footnotesize
	\caption{The extracted values of the QGP temperature (in MeV) for different centrality regions of Ru+Ru and Zr+Zr collisions at $\sqrt{s_\mathrm{NN}} = 200$~GeV, compared between using heavy quark decayed dielectron spectra with different model setups.}
	\label{tab:200_0.0003}
	\renewcommand{\arraystretch}{1.15}
	\setlength{\tabcolsep}{3.5pt}
	\vspace{5pt}
	\begin{tabular}{ccccccc}
		\hline
		centrality 
		& vac. frag. 
		& med.+frag. 
		& med.+frag.
		& med.+frag.+coal.
		& med.+frag.+coal. \\
		&  &  & (w/o $b$ subtr.) &  & (w/o $b$ subtr.) \\
		\hline
		0-10\% & $271 \pm 22$ & $313 \pm 20$ & $327 \pm 20$ & $306 \pm 19$ & $311 \pm 19$ \\
		10-40\% & $313 \pm 15$ & $351 \pm 15$ & $367 \pm 15$ & $340 \pm 13$ & $346 \pm 13$ \\
		0-80\% & $295 \pm 11$ & $330 \pm 11$ & $345 \pm 11$ & $320 \pm 10$ & $327 \pm 10$ \\
		40-80\% & $333 \pm 18$ & $357 \pm 18$ & $372 \pm 18$ & $345 \pm 17$ & $353 \pm 17$ \\
		\hline
	\end{tabular}
	\vspace{2mm}
\end{table*}

In Tab.~\ref{tab:54.4_0.0003}, we present the extracted values of the QGP temperature in different centrality regions of Au+Au collisions at $\sqrt{s_\mathrm{NN}} = 54.4$~GeV, all obtained using $\epsilon_c = 0.0003$ for the Peterson fragmentation function and the model parameters of heavy-quark-QGP interactions listed in Sec.~\ref{sec:2}. Conclusions similar to those for the 0-80\% centrality in Fig.~\ref{fig:54.4} can be drawn by comparing between different model setups: heavy quark energy loss softens the charm decayed dielectron spectra, leading to a larger extracted QGP temperature, while coalescence in hadronization hardens the spectra, reducing the extracted temperature. Overall, by implementing our full model calculation, higher values of temperature are expected compared to applying the vacuum baselines of the charm decayed dielectron spectra in the cocktail sums.

The similar analysis is also extended to Isobar collisions (Ru+Ru and Zr+Zr) at $\sqrt{s_\mathrm{NN}}=200$~GeV, as shown in Fig.~\ref{fig:200}. At this collision energy, contributions from bottom quark decay may also be sizable. In Fig.~\ref{fig:200}(a), we present different model calculations of dielectron spectra from both charm and bottom quark decays in 0-10\% Isobar collisions. The $\epsilon_Q$ parameter in the Peterson fragmentation function is set as 0.0003 for charm quarks as before. We have verified that the invariant mass spectrum of bottom decayed dielectrons is not sensitive to this parameter, and therefore keep the default value of 0.005 for bottom quarks in Pythia. Since we do not implement acceptance corrections as experimental analysis does, we need to normalize our vacuum baseline of bottom decayed dielectrons to that obtained by the STAR simulation. As discussed earlier, these corrections have little impact on the spectrum of charm decayed electrons in the intermediate mass region. With these settings, our vacuum baselines of charm (blue line) and bottom decayed dielectrons (dark blue line) are both consistent with those applied by the STAR Collaboration in their data analysis (green and light blue lines). Compared to the baselines, we see that charm quark energy loss softens the spectrum of charm decayed dielectrons (gold line),  while coalescence hardens the spectrum (red line), same as what we found in Au+Au collisions at $\sqrt{s_\mathrm{NN}}=54.4$~GeV. On the other hand, we observe little medium modification of the bottom decayed dielectron spectrum when pure fragmentation is applied on medium-modified bottom quarks (pink line). This is due to both the small energy loss experienced by bottom quarks inside the QGP and the flat invariant mass spectrum of the bottom decayed dielectrons. Since only dielectrons decayed from $B$ mesons and $\Lambda_b$ baryons are included in this analysis, but bottom quarks can also turn into heavier baryons beyond $\Lambda_b$, with branching ratios larger in A$+$A collisions via coalescence than in $p+p$ collisions via pure fragmentation~\cite{Cao:2019iqs}, here we observe smaller yield of bottom decayed dielectrons via hybrid fragmentation-coalescence hadronization (purple line) than via pure fragmentation (pink line).

Using the dielectron spectra from charm and bottom quark decays, we recalculate the invariant mass spectra of thermal dielectron pairs and extract the QGP temperature in the intermediate mass region ($1.0 < M_{ee} < 2.8~\mathrm{GeV}$), as shown in Fig.~\ref{fig:200}(b). The extracted values are labeled in the legend. Our value extracted based on the vacuum baseline of heavy quark decayed electrons ($271 \pm 22$~MeV) is consistent with that obtained by STAR ($270 \pm 21$~MeV). Using dielectrons produced by medium-modified heavy quarks followed by pure fragmentation, this value rises to $313 \pm 20$~MeV. After coalescence is included for hadronization, it drops to $306 \pm 19$~MeV. Therefore, one can conclude that in 0-10\% Isobar collisions, heavy quark energy loss can contribute to about 42~MeV systematic uncertainty in the extracted temperature of the QGP, and hadronization can contribute to about 7~MeV uncertainty.

The extracted temperatures for different centralities of Isobar collisions are presented in Tab.~\ref{tab:200_0.0003}, all obtained using $\epsilon_c = 0.0003$ and $\epsilon_b = 0.005$ for the Peterson fragmentation function, and the model parameters of heavy-quark-QGP interactions listed in Sec.~\ref{sec:2}. Again, one can observe that compared to using the vacuum baselines of heavy quark decayed dielectrons in analysis, including in-medium scatterings of heavy quarks followed by pure fragmentation leads to larger values of the extracted temperature, while further including coalescence in hadronization mildly reduces these values. Overall, our full model calculation suggests that the nuclear modification of heavy flavor hadrons, including both heavy quark energy loss and their in-medium hadronization, can result in 12 to 35~MeV (depending on centrality) larger value of the extracted temperature compared to using the in-vacuum spectra of heavy quark decayed dielectrons in analyzing the experimental data. Note that unlike for Au+Au collisions at $\sqrt{s_\mathrm{NN}}=54.4$~GeV where the uncertainties in the fitted temperature are large, the systematic uncertainties from medium modification of heavy flavor hadrons here are comparable to or even larger than the uncertainties from data fitting for Isobar collisions at $\sqrt{s_\mathrm{NN}}=200$~GeV. This medium modification would become even more important for heavy-ion collisions with larger colliding nuclei or higher beam energies, like Au+Au collisions at $\sqrt{s_\mathrm{NN}}=200$~GeV and Pb+Pb collisions at the LHC energies, because of the increase of both heavy quark yield and their energy loss in larger QGP systems. Therefore, one should no longer ignore this source of uncertainty for precisely constraining the QGP temperature in the future.

In Tab.~\ref{tab:200_0.0003}, we also estimate the systematic uncertainty from whether the bottom quark contribution is included. When pure fragmentation is applied to medium-modified heavy quarks, about 14 to 16~MeV higher value would be obtained if the bottom quark contribution is ignored, this uncertainty is reduced to 5 to 8~MeV when coalescence is included in hadronization. Therefore, decay of bottom quarks does affect the temperature extraction, but its impact is smaller than the current uncertainties of the extracted values.

In this section, in order to reproduce the vacuum baselines of the invariant mass spectra of heavy quark decayed dielectrons used by the STAR Collaboration, we adjust the fragmentation parameter to $\epsilon_c = 0.0003$ for charm quarks. We have verified that if the default value in Pythia ($\epsilon_c = 0.05$) is used (which does not reproduce the vacuum baseline used by the STAR analysis), effects of heavy quark energy loss would become smaller, and so would the related uncertainties of the extracted temperatures. One may understand the fragmentation process as a kind of ``effective energy loss" of a heavy quark when it turns into a hadron. Therefore, when a much softer fragmentation function ($\epsilon_c = 0.05$) is used, the stronger ``energy loss" of heavy quarks during hadronization could overshadow their energy loss inside the QGP and weaken effects of the latter. In this case, when our full calculation is implemented, the systematic uncertainty due to medium modification of heavy flavor hadrons is negligible for Au+Au collisions at $\sqrt{s_\mathrm{NN}}=54.4$~GeV compared to the errors of the current data, and is about 6-12~MeV for Isobar collisions at $\sqrt{s_\mathrm{NN}}=200$~GeV. We expect even when using the Peterson fragmentation function with its default parameters, the systematic uncertainty arising from heavy quark interactions with the QGP cannot be neglected in extracting the QGP temperature at higher collision energies in the future. Alternatively, we validate the robustness of our conclusion by refitting the $R_\mathrm{AA}$ of medium-to-high-$p_\mathrm{T}$ $D$ mesons at RHIC and the LHC, readjusting $\alpha_\mathrm{s}$ and $\sigma$ in Eq.~(\ref{eq:potential}) when $\epsilon_c=0.0003$ is used. With the readjusted model parameters, the invariant mass spectra of dielectrons are recalculated and the values of the QGP temperature are re-extracted. We find that the re-extracted central values of the QGP temperature are only 2 to 5~MeV lower than those listed in Tabs.~\ref{tab:54.4_0.0003} and~\ref{tab:200_0.0003}. Our conclusions regarding the effects of the medium modification of heavy flavor hadrons on the extracted QGP temperature remain unchanged.

\begin{table}[H]
	\centering
	\footnotesize
	\begin{threeparttable}
		\caption{Average temperature values directly extracted from the hydrodynamic profiles of the QGP.}
		\label{tab:avgT_hydro}
		\setlength{\tabcolsep}{6pt}
		\renewcommand{\arraystretch}{1.1}
		\begin{tabular}{ccc}
			\toprule
			Centrality & Au+Au 54.4 GeV & Ru+Ru / Zr+Zr 200 GeV \\
			\midrule
			0-10\%   & 223 & 236 \\
			10-40\%  & 213 & 223 \\
			0-80\%   & 203 & 209 \\
			40-80\%  & 193 & 200 \\
			\bottomrule
		\end{tabular}
	\end{threeparttable}
\end{table}

This work focuses on the impact of heavy quark energy loss and hadronization on the extraction of the QGP temperature. Nevertheless, the extracted temperature may also be influenced by other effects. In peripheral A$+$A collisions, event selection and geometric biases can become significant, potentially leading to an overestimation of the heavy quark yield and thus the impact of their energy loss~\cite{Loizides:2017sqq,Jonas:2021xju}. When these biases are properly taken into account, the uncertainty in the extracted temperature due to heavy-quark-QGP interactions should be reduced in peripheral collisions. In addition, cold nuclear matter (CNM) effects are not considered in this work. They are expected to suppress the low $M_{ee}$ and enhance the high $M_{ee}$ charm decayed dielectron productions, partially offsetting the effect of heavy quark energy loss in the QGP. As a result, the extracted QGP temperature should also move closer to that obtained using heavy quark spectra without considering their energy loss. Detailed studies on these additional effects beyond heavy quark energy loss and hadronization will be left for future exploration.

\begin{figure*}[tbp]
	\centering
	\includegraphics[width=0.72\textwidth]{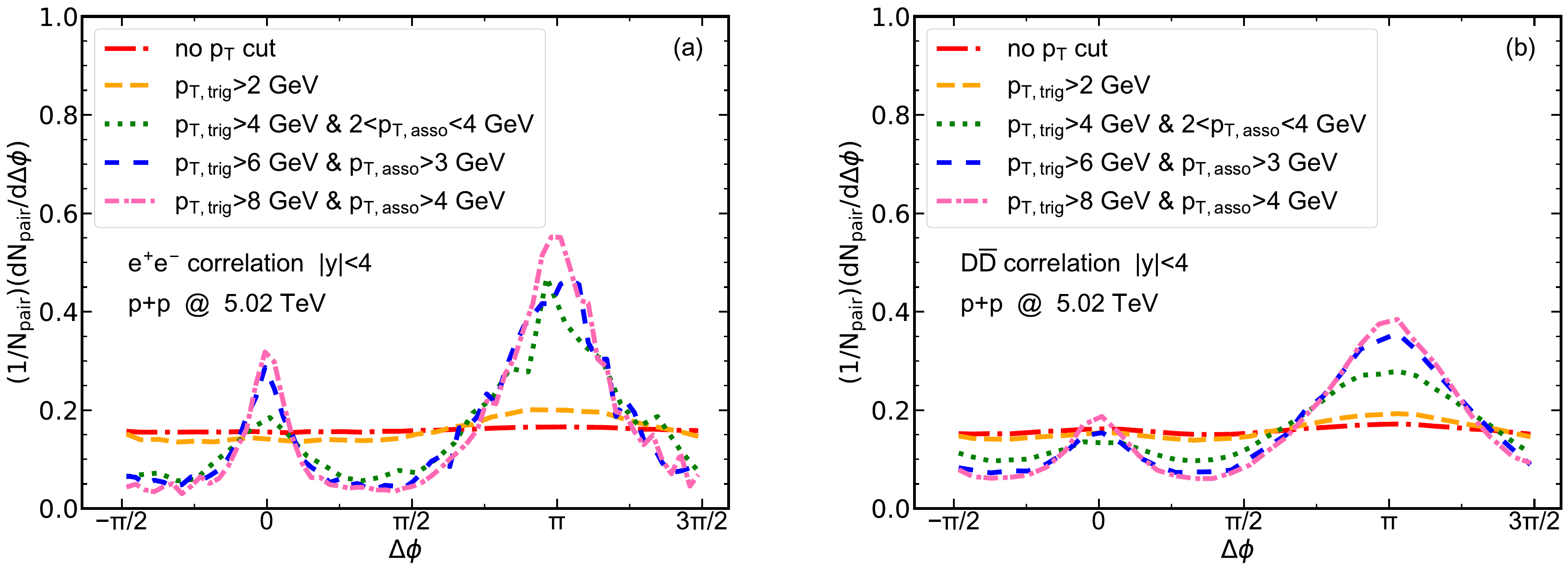}
	\caption{(Color online) Angular correlations between (a) $e^{+}e^{-}$ pairs and (b) $D\overline{D}$ pairs in $p+p$ collisions at $\sqrt{s} = 5.02$~TeV, compared between different $p_\mathrm{T}$ cuts on the final state particles.}
	\label{fig:pp}
\end{figure*}

Last but not least, we present the average temperatures of our hydrodynamic media in Tab.~\ref{tab:avgT_hydro}, which can be compared to those extracted from the dilepton spectra in Tab.~\ref{tab:54.4_0.0003} and Tab.~\ref{tab:200_0.0003}. For each hydrodynamic profile, we average the temperatures over its cells at mid rapidity using the local energy density as a weight factor. Local temperature of each cell is required to be above $T_{\mathrm{pc}}$ for its contribution to the global average. The average temperatures we obtain are consistent with those presented in Ref.~\cite{Churchill:2023zkk}. As shown in Ref.~\cite{Churchill:2023zkk}, these average temperatures should be consistent with the effective temperatures extracted from thermal dileptons. Nevertheless, we see the values listed in Tab.~\ref{tab:avgT_hydro} are lower than those listed in Tab.~\ref{tab:54.4_0.0003} and Tab.~\ref{tab:200_0.0003} (labelled with ``med.+frag.+coal."). This could result from the contributions from dileptons produced in the pre-equilibrium stage of heavy-ion collisions and the contamination from multi-pion decays~\cite{vanHees:2007th,Endres:2014zua} which are hard to be subtracted from the dilepton spectra in experiments.

\begin{figure*}[tbp]
	\centering
	\includegraphics[width=0.72\textwidth]{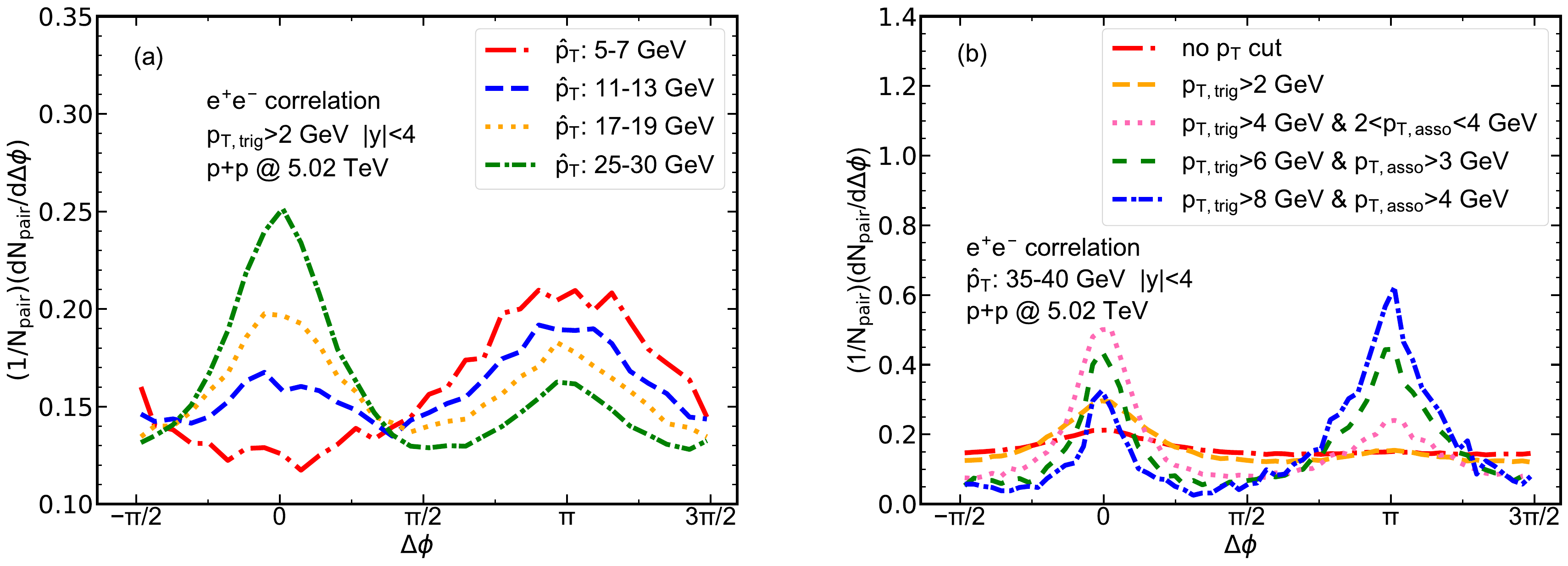}
	\caption{(Color online) Angular correlations between $e^{+}e^{-}$ pairs in $p+p$ collisions at $\sqrt{s} = 5.02$~TeV, compared between (a) different hard scattering scales with the final $p_\mathrm{T}$ cut fixed, and (b) different final $p_\mathrm{T}$ cuts with the hard scattering scale fixed.}
	\label{fig:pp_bin}
\end{figure*}

\section{Angular correlations of dielectrons from charm hadron decays}
\label{sec:4}

In this section, we investigate the nuclear modification of the angular correlation between dielectron pairs and its dependence on the interaction dynamics between heavy quarks and the QGP. As a baseline, we start with the correlation between $e^{+}e^{-}$ pairs in $p+p$ collisions at a center-of-mass energy of $\sqrt{s}=5.02$~TeV. The correlation function is defined as the normalized distribution of $e^{+}e^{-}$ pairs with respect to their relative angles: $(1/N^{e^{+}e^{-}})(dN^{e^{+}e^{-}}/d\Delta\phi)$. In Fig.~\ref{fig:pp}(a), we observe this correlation function strongly relies on the kinematic region of $e^{+}e^{-}$ pairs: its $\Delta\phi$ dependence is weak when no transverse momentum ($p_\mathrm{T}$) cut is implemented on the final state electrons, but exhibits more prominent peak structures on both near ($\Delta\phi \sim 0$) and away ($\Delta\phi \sim \pi$) sides as higher $p_\mathrm{T}$ cuts are implemented on the leading (trigger) and its associated electrons (or positrons).
The away side peak arises from the back-to-back production of the initial $c\bar{c}$ pairs in the transverse plane via the $q\bar{q} \to c\bar{c}$ and $gg \to c\bar{c}$ processes. 
With a higher $p_\mathrm{T}$ cut, the selected electrons originate from more energetic charm quarks produced in harder initial scatterings, and thus experience weaker deflection by parton shower, hadronization and decay, and preserve more distinct back-to-back configurations.  
The near side peak is mainly attributed to the splittings of gluons produced from the initial hard scatterings. One can verify that this peak is significantly suppressed when only $q\bar{q} \to c\bar{c}$ and $gg \to c\bar{c}$ are switched on for charm quark production in Pythia, although for scatterings with very large transverse momentum exchange ($\hat{p}_\mathrm{T}$, or the ``pTHat" parameter in Pythia), high-energy charm quarks may emit gluons that further split into collinear $c\bar{c}$ pairs.
A higher final $p_\mathrm{T}$ cut on electrons triggers scatterings with higher initial $\hat{p}_\mathrm{T}$, which produce more energetic gluons that are easier to split into $c\bar{c}$ pairs, and thus enhances the near side peak. In Fig.~\ref{fig:pp}(b), the angular correlation between $D\overline{D}$ is shown for comparison. Considering the energy shift in the $D\to e$ decay process, the same final $p_\mathrm{T}$ cut implies higher initial $\hat{p}_\mathrm{T}$ for $e^{+}e^{-}$ than $D\overline{D}$ productions. Therefore, with the same $p_\mathrm{T}$ cut, the $e^{+}e^{-}$ correlation shows stronger peak patterns than the $D\overline{D}$ correlation. 

To better understand the origins of the two peaks, we further explore their dependences on the initial hard scattering ($\hat{p}_\mathrm{T}$) scale and the final kinematic ($p_\mathrm{T}$) scale. In Fig.~\ref{fig:pp_bin}(a), we fix the $p_{\mathrm{T,trig}}>2$~GeV cut for $e^{+}e^{-}$ pairs and show their angular correlations for different $\hat{p}_\mathrm{T}$ bins. At low $\hat{p}_\mathrm{T}$, the LO processes ($gg \to c\bar{c}$ and $q\bar{q} \to c\bar{c}$) dominate, resulting in the more prominent away side peak than the near side peak. However, at high $\hat{p}_\mathrm{T}$, it becomes easier for energetic gluons to split into soft $c\bar{c}$ pairs. Due to the much larger cross section of gluon production than charm quark production at the LO, the near side peak becomes more prominent than the away side peak in the $e^{+}e^{-}$ correlation. Since charm quarks produced from gluon splittings are usually softer than those directly produced from the LO processes, the near side peak can be suppressed as one applies sufficiently high $p_\mathrm{T}$ cuts on the final state particles. This is verified in Fig.~\ref{fig:pp_bin}(b), where we fix $\hat{p}_\mathrm{T}\in (35, 40)$~GeV and vary the $p_\mathrm{T}$ cuts on the final state electrons. Another factor influencing the angular correlation function is the uncorrelated $c\bar{c}$ pairs. As $\hat{p}_\mathrm{T}$ increases, it is easier to obtain multiple $c\bar{c}$ pairs in each $p+p$ collision event. The uncorrelated $c\bar{c}$ background can broaden the peak structures of the normalized correlation function. Effects of this uncorrelated background can be reduced by implementing sizable $p_\mathrm{T}$ cuts on the final state electrons. Similar conclusions can also be drawn for correlation between $D\overline{D}$ pairs except that they are less sensitive to the final state $p_\mathrm{T}$ cuts than $e^{+}e^{-}$ pairs.

\begin{figure*}[tbp]
	\centering
	\includegraphics[width=0.92\textwidth]{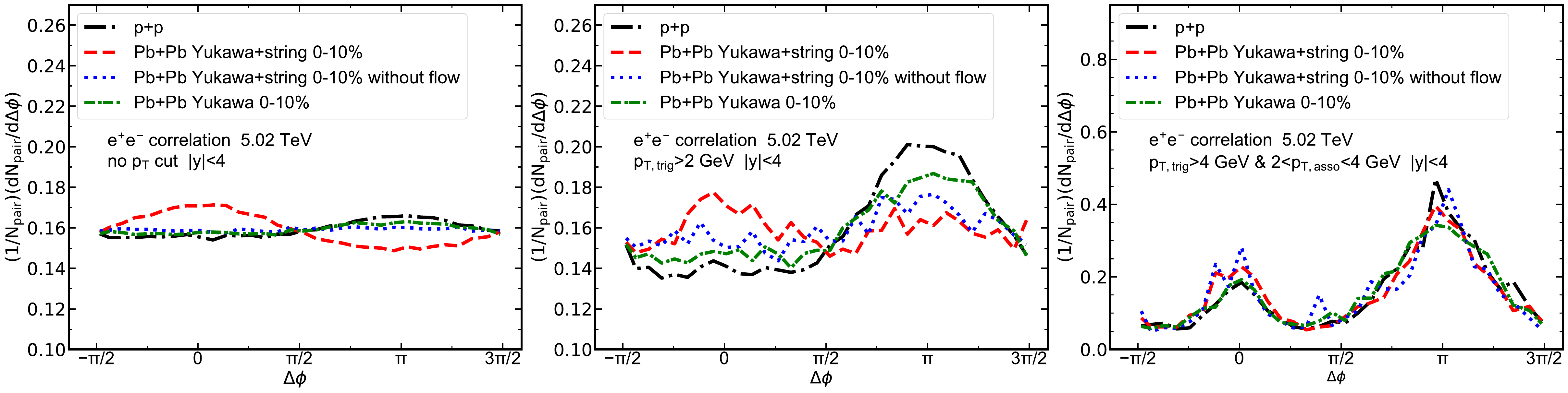}
	\caption{(Color online) Angular correlations between $e^{+}e^{-}$ pairs in central (0-10\%) Pb+Pb collisions at $\sqrt{s_\mathrm{NN}} = 5.02$~TeV, compared between with and without the QGP flow effect, Yukawa$+$string and Yukawa interactions, and the $p+p$ baselines. Different panels are for different $p_\mathrm{T}$ cuts on the final state electrons.}
	\label{fig:AA_y+s}
\end{figure*}

\begin{figure*}[tbp]
	\centering
	\includegraphics[width=0.92\textwidth]{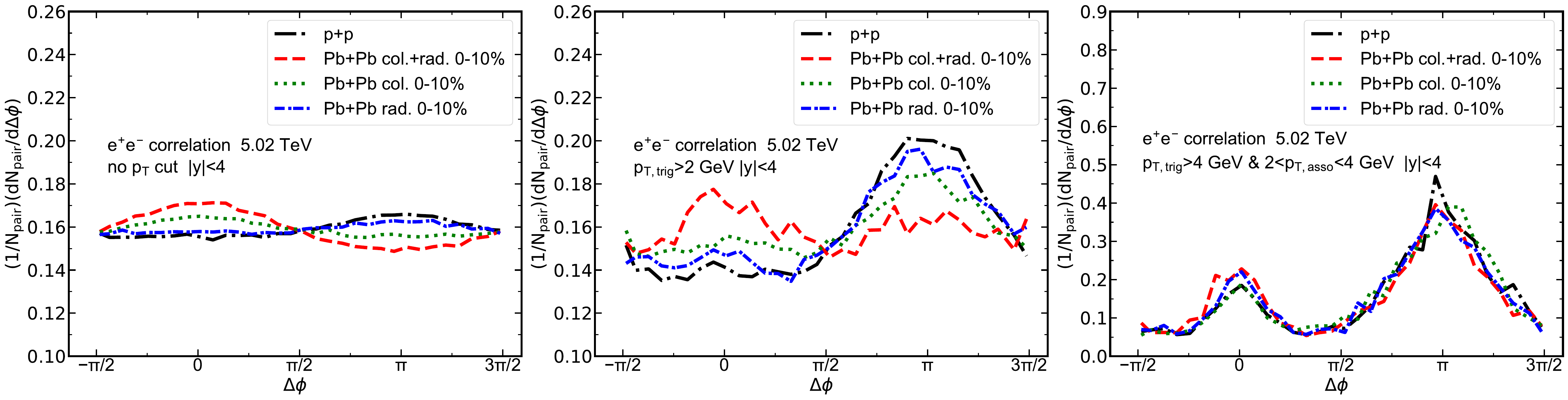}
	\caption{(Color online) Angular correlations between $e^{+}e^{-}$ pairs in central (0-10\%) Pb+Pb collisions at $\sqrt{s_\mathrm{NN}} = 5.02$~TeV, compared between contributions from elastic and inelastic scatterings, their combination, and the $p+p$ baselines. Different panels are for different $p_\mathrm{T}$ cuts on the final state electrons.}
	\label{fig:AA_col+rad}
\end{figure*}

It would be interesting if one could identify a kinematic region that is dominated by the $g \to c\bar{c}$ process and investigate its medium modification in heavy-ion collisions. However, by convoluting the correlation functions at different $\hat{p}_\mathrm{T}$'s with the steeply falling cross section as $\hat{p}_\mathrm{T}$ increases, the final state $e^+e^-$ correlations always show more prominent peaks at the away side than the near side. One possible way to isolate the $g \to c\bar{c}$ process might be triggering a hadron, instead of a heavy quark decay product, and study an $e^+e^-$ pair on its backward direction. This will be left for a future exploration.

In Fig.~\ref{fig:AA_y+s}, we study the nuclear modification of the angular correlation of $e^+e^-$ pairs in central (0-10\%) Pb+Pb collisions at  $\sqrt{s_\mathrm{NN}} = 5.02$ TeV. Here, we do not include correlations between $e^+$ and $e^-$ produced by different nucleon-nucleon collisions in a nucleus-nucleus collision. This uncorrelated background can be assumed uniformly distributed with respect to $\Delta \phi$ and subtracted in experimental measurements with an estimation of the average binary collision number of a nucleus-nucleus collision within a given centrality region. On the other hand, contributions from uncorrelated $e^+e^-$ pairs in each nucleon-nucleon collision are taken into account.

When both Yukawa and string interactions are included for charm quark scatterings with the QGP, considerable nuclear modification effects can be observed when there is no $p_\mathrm{T}$ cut on the final state electrons, or the cuts are not high (the left two panels of Fig.~\ref{fig:AA_y+s}). Compared to the $p+p$ baselines where the away side peaks dominate, we see prominent near side peaks emerge in Pb+Pb collisions instead. This results from the radial flow of the QGP, which drives the low $p_\mathrm{T}$ $c\bar{c}$ pairs from back-to-back to collinear configuration. Meanwhile, this suggests low $p_\mathrm{T}$ charm quarks can quickly approach thermalization with the medium and flow with the medium. This phenomenon was referred to as the ``partonic wind effect" in literature~\cite{Zhu:2007ne}, and is expected to enhance the $J/\psi$ regeneration in heavy-ion collisions. As shown in the left two panels, when the coupling between heavy quarks and the flow velocity of the QGP is switched off in the LBT simulation, the near side peaks are significantly reduced. Broadening of the away side peaks, due to scatterings between heavy quarks and the QGP, still exists when the flow velocity is switched off. Moreover, within our model, motions of low $p_\mathrm{T}$ heavy quarks are mainly driven by the string interaction. As a result, medium effects become much weaker when only Yukawa interaction is taken into account. With high $p_\mathrm{T}$ cuts (the right panel of Fig.~\ref{fig:AA_y+s}), nuclear modifications of the angular correlation of $e^+e^-$ pairs appear relatively weak because energetic charm quarks can preserve their directions well when ploughing through the medium. 

In Fig.~\ref{fig:AA_col+rad}, we investigate effects of elastic and inelastic energy loss of heavy quarks on the angular correlations between $e^+e^-$ pairs. One can observe that inelastic scatterings, or gluon emissions, alone have little impact on modifying their angular correlations. This is because the gluon emission process is dominated by collinear splitting within the higher twist formalism~\cite{Zhang:2003wk,Guo:2000nz,Majumder:2009ge} we apply. Contrarily, elastic scatterings significantly deflect the heavy quark trajectories, leading to the emergence of the near side peaks and the broadening of the away side peaks when no or moderate $p_\mathrm{T}$ cuts are implemented (the left two panels). Note that although inelastic scatterings do not alter the direction of heavy quark motion directly, they contribute to considerable energy loss of heavy quarks, and therefore expedite their deflections driven by elastic scatterings. These findings qualitatively agree with an earlier study on the angular correlations between $D\overline{D}$ pairs~\cite{Cao:2015cba}. However, unlike the previous work, we no longer fit our model parameters to the heavy flavor data when elastic or inelastic scattering is applied alone, since neither of them can quantitatively describe the data individually. The model parameters are determined by the $R_\mathrm{AA}$ and $v_2$ data of $D$ mesons when elastic and inelastic processes are combined, and their separate contributions to the nuclear modification of $e^+e^-$ correlation are presented here. It should be noted that only the results from the combined elastic and inelastic scatterings represent the physical predictions to be compared with experimental data in the future.

\begin{figure*}[tbp]
	\centering
	\includegraphics[width=0.95\textwidth]{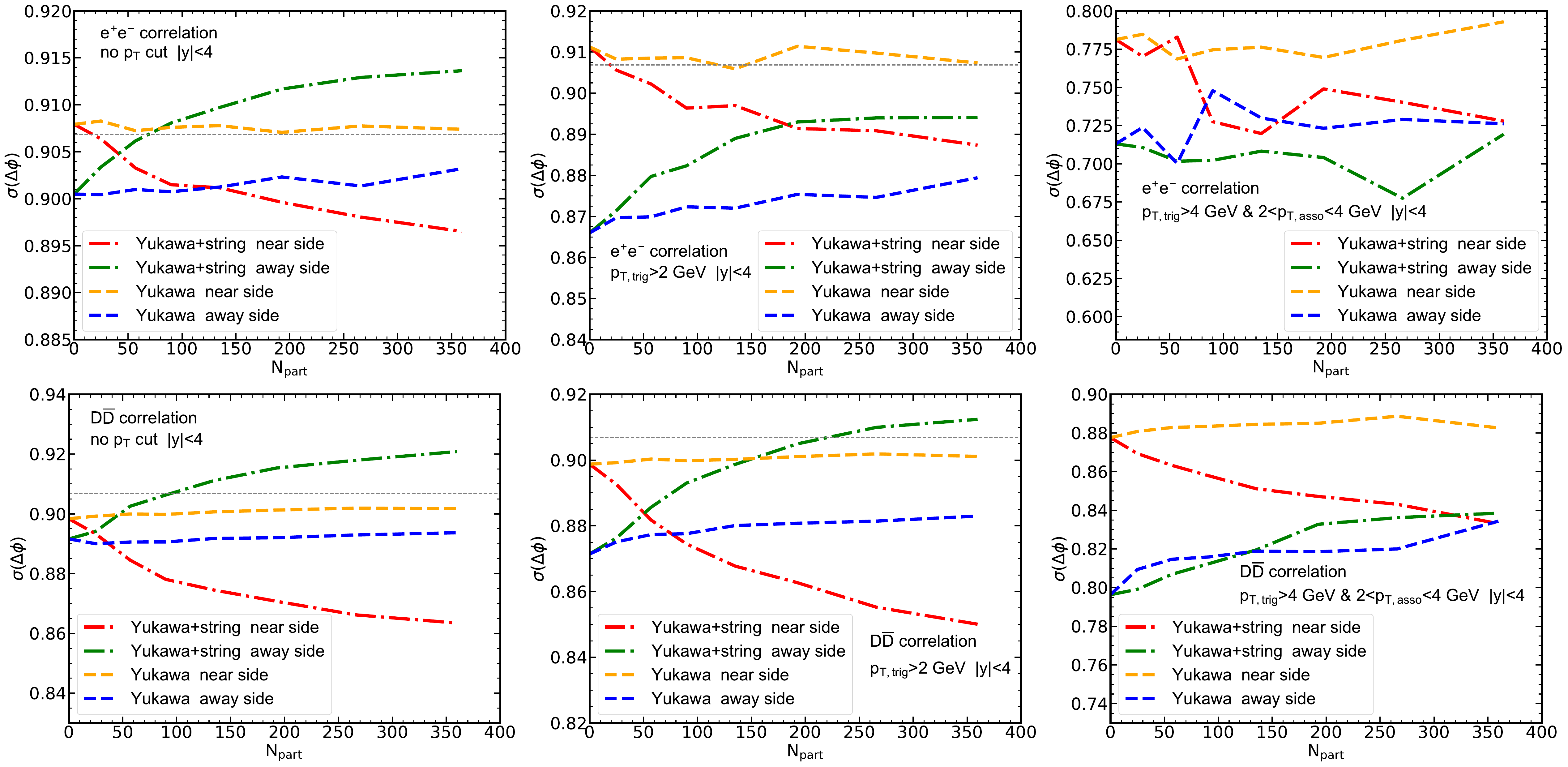}
	\caption{(Color online) The participant number dependence of the variance of the angular correlation functions of $e^+e^-$ pairs (upper panels) and $D\overline{D}$ pairs (lower panels) at near and away sides in Pb+Pb collisions at $\sqrt{s_\mathrm{NN}} = 5.02$~TeV, compared between different scattering mechanisms and different kinematic cuts.}
	\label{fig:sigma}
\end{figure*}

To quantify the medium modification of angular correlations, we evaluate the variance of the $\Delta\phi$ distributions at both near side ($-\pi/2$, $\pi/2$) and away side ($\pi/2$, $3\pi/2$). Within a given $\Delta\phi$ region, the variance is defined as $\sigma=\sqrt{(\Delta\phi-\langle\Delta\phi\rangle)^2}=\sqrt{\langle\Delta\phi^2\rangle-\langle\Delta\phi\rangle^2}$, with $\langle \ldots \rangle$ denoting averaging. Shown in Fig.~\ref{fig:sigma} is the participant number ($N_\mathrm{part}$) dependence of $\sigma$ for both $e^+e^-$ correlations (upper panels) and $D\overline{D}$ correlations (lower panels) in Pb+Pb collisions at $\sqrt{s_\mathrm{NN}} = 5.02$ TeV, with $N_\mathrm{part}=0$ corresponding to their $p+p$ baselines. In the left four panels, a baseline of $\pi/\sqrt{12}$ is also shown for comparison, which is the variance of a uniform distribution of $\Delta\phi$ within $(-\pi/2, \pi/2)$ or $(\pi/2, 3\pi/2)$. Therefore, $\sigma<\pi/\sqrt{12}$ indicates a peak structure, while $\sigma>\pi/\sqrt{12}$ indicates a dip structure.

When both Yukawa and string interactions are included, the near side variance decreases with increasing $N_\mathrm{part}$, indicating that the angular correlation around $\Delta\phi = 0$ evolves from a dip or a weak peak to a more pronounced peak. This behavior can be attributed to the stronger radial flow of the QGP in more central collisions that drives a pair of $c\bar{c}$ quarks towards the same direction. The away side variance increases with $N_\mathrm{part}$, indicating the peak structure around $\Delta\phi = \pi$ gets broadened or even turns into a dip structure when charm quark scattering with the QGP becomes stronger in more central collisions. In the RHIC environment, one may observe a non-monotonic dependence of $\sigma$ on $N_\mathrm{part}$ on the away side with certain $p_\mathrm{T}$ cuts implemented~\cite{Cao:2015cba}, which results from the competition between trigger bias and momentum broadening. The former selects $c\bar{c}$ pairs produced from higher $\hat{p}_\mathrm{T}$ with narrower away side peak, while the latter broadens the peak. However, due to the harder heavy quark spectra, and the less pronounced away side peak structure of $c\bar{c}$ pairs at the LHC energy than at the RHIC energy~\cite{Zhu:2007ne}, in this work, the effect of trigger bias is weak and thus $\sigma$ monotonically increases with $N_\mathrm{part}$ on the away side within the kinematic regions we explore for the LHC environment. When only the Yukawa interaction is included, the variance of the near side correlation shows little dependence on $N_\mathrm{part}$, whereas that of the away side exhibits a slight increase with $N_\mathrm{part}$. This confirms again that the string interaction dominates the deflection of heavy quarks within our model.

\section{Summary}
\label{sec:5}

We conduct a systematical study on the heavy flavor decayed dileptons in relativistic heavy-ion collisions, including nuclear modification of both their invariant mass spectra and their angular correlation functions. Charm and bottom quark productions in the initial hard nucleon-nucleon scatterings are simulated by Pythia. Interactions between heavy quarks and the QGP are simulated using the LBT model, which includes both elastic and inelastic scatterings, and both Yukawa and string types of interactions. Medium-modified heavy quarks then turn into heavy flavor hadrons via a hybrid fragmentation-coalescence model, and further decay into electrons via Pythia. 

Within this framework, we find that heavy quark energy loss inside the QGP softens the invariant mass spectrum of heavy quark decayed dielectrons, and coalescence hardens the spectrum. Since this spectrum serves as a crucial background that needs to be subtracted for obtaining the thermal dielectron spectrum, the nuclear modification of heavy flavor hadrons affects the value of the QGP temperature extracted from the thermal spectrum. Compared to using Pythia alone to simulate heavy quark decayed dielectrons, including heavy quark energy loss leads to a higher value of temperature extracted from the dielectron spectrum, and including coalescence leads to a lower value. Our full calculation suggests a larger value of temperature than applying vacuum spectrum of heavy quark decayed dielectrons in the analysis. While this systematic uncertainty is smaller than the fitting errors of the QGP temperature in Au+Au collisions at $\sqrt{s_\mathrm{NN}}=54.4$~GeV, it can reach 12 to 35~MeV (depending on centrality) in Isobar collisions at $\sqrt{s_\mathrm{NN}}=200$~GeV, comparable to or even larger than the corresponding fitting errors. It should be noted that in order to reproduce the invariant mass spectra of dielectrons used in the STAR analysis, we temporarily adjust the $\epsilon_c$ parameter in the Pythia fragmentation function of charm quarks from its default value of 0.05 to 0.0003 in Sec.~\ref{sec:3}. Although the default value was used in our earlier study~\cite{Xing:2021xwc} where our model parameters were tuned, we have verified that the qualitative conclusions of this work are not affected by this inconsistency. Using $\epsilon_c=0.05$ with our original model parameters or using $\epsilon_c=0.0003$ with model parameters refit to the medium-to-high-$p_\mathrm{T}$ $D$ meson $R_\mathrm{AA}$ at RHIC and the LHC can both reduce the effects of medium modification of heavy flavor hadrons on the extracted QGP temperature. However, the remaining effects are still non-negligible for measurements in Isobar collisions at $\sqrt{s_\mathrm{NN}}=200$~GeV. Therefore, a rigorous treatment of the nuclear modification of heavy flavor hadrons is necessary for precisely probing the QGP temperature in the future, especially for heavy-ion collisions with larger colliding nuclei or at higher beam energies.

The angular correlations between heavy quark decayed $e^+e^-$ pairs result from complicated competitions between different initial production processes of heavy quarks, and different interaction mechanisms between heavy quarks and the QGP. Pair productions of charm quarks, $q\bar{q} \to c\bar{c}$ and $gg \to c\bar{c}$, generate away side peaks of the $e^+e^-$ correlations, while gluon splittings, $g \to c\bar{c}$, generate near side peaks. Their significance depends on both the initial hard scattering scale ($\hat{p}_\mathrm{T}$) and the final kinematic ($p_\mathrm{T}$) range of electrons. Inside the QGP, the radial flow of the medium can drive low $p_\mathrm{T}$ $c\bar{c}$ pairs towards the same direction and leads to a prominent near side peak of $e^+e^-$ correlation when the final $p_\mathrm{T}$ cut is not high. Meanwhile, scatterings between charm quarks and the QGP broaden the away side peak. In Pb+Pb collisions at $\sqrt{s_\mathrm{NN}}=5.02$~TeV, the increasing radial flow from peripheral to central collisions leads to a monotonic decrease of the near side variance of the angular distribution as $N_\mathrm{part}$ increases, while momentum broadening leads to a monotonic increase of the away side variance. Within our model, nuclear modification of angular correlations is dominated by elastic scatterings and string type of interactions rather than inelastic scatterings and Yukawa type of interactions. These conclusions also apply to correlations between $D\overline{D}$ pairs. Future measurements on these correlations would provide deeper insights into the interaction dynamics of heavy quarks with the QGP.

\Acknowledgements{We are grateful for the generous help from Qian Yang in setting up comparisons between model calculations and experimental data, and valuable discussions with Zhen Wang, Xianwen Bao, Jiaxuan Luo, Zebo Tang, Zaochen Ye, Xiaowen Li, and Xiang-Yu Wu. This work is supported by the National Natural Science Foundation of China (NSFC) under Grant Nos.~12575146, 12175122, 2021-867, 12505154, 12225503, and~12321005.}

\InterestConflict{The authors declare that they have no conflict of interest.}

\end{multicols}
\end{document}